# Effects of Recording Condition and Number of Monitored Days on Discriminative Power of the Daily Phonotrauma Index

Hamzeh Ghasemzadeh [a,b,c], Robert E. Hillman [a,b,d,e], Jarrad H. Van Stan [a,b,e], Daryush D. Mehta [a,b,d,e]

[a] Center for Laryngeal Surgery and Voice Rehabilitation, Massachusetts General Hospital, Boston, MA, USA
[b] Department of Surgery, Harvard Medical School, Boston, MA, USA
[c] Department of Communicative Sciences and Disorders, Michigan State University, East Lansing, MI, USA
[d] Speech and Hearing Bioscience and Technology, Division of Medical Sciences, Harvard Medical School, Boston, MA, USA
[e] MGH Institute of Health Professions, Boston, MA, USA

Running Title: Effects of Recording Condition and Duration on DPI

Corresponding Author:
Hamzeh Ghasemzadeh
Center for Laryngeal Surgery and Voice Rehabilitation
Massachusetts General Hospital, One Bowdoin Square, 11th Floor, Boston, MA 02114
Email: hghasemzadeh@mgh.harvard.edu

Disclosure:
    Drs. Robert Hillman and Daryush Mehta have a financial interest in InnoVoyce LLC, a company focused on developing and commercializing technologies for the prevention, diagnosis, and treatment of voice-related disorders. Dr. Hillman's and Dr. Mehta's interests were reviewed and are managed by Massachusetts General Hospital and Mass General Brigham in accordance with their conflict-of-interest policies.

**Abstract:**

**Objective**: The Daily Phonotrauma Index (DPI) can quantify pathophysiological mechanisms associated with daily voice use in individuals with phonotraumatic vocal hyperfunction (PVH). Since DPI was developed based on week-long ambulatory voice monitoring, this study investigated if DPI can achieve comparable performance using (1) short laboratory speech tasks and (2) fewer than seven days of ambulatory data.

**Method**: An ambulatory voice monitoring system recorded the vocal function/behavior of 134 females with PVH and vocally healthy matched controls in two different conditions. In the lab, the participants read the first paragraph of the Rainbow Passage and produced spontaneous speech (in-lab data). They were then monitored for seven days (in-field data). Separate DPI models were trained from in-lab and in-field data using the standard deviation of the difference between the magnitude of the first two harmonics (H1-H2) and the skewness of neck-surface acceleration magnitude. First, 10-fold cross-validation evaluated classification performance of in-lab and in-field DPIs. Second, the effect of the number of ambulatory monitoring days on the accuracy of in-field DPI classification was quantified.

**Results**: The average in-lab DPI accuracy computed from the Rainbow passage and spontaneous speech were, respectively, 57.9% and 48.9%, which are close to chance performance. The average classification accuracy of in-field DPI was significantly higher with a very large effect size (73.4%, Cohen's D = 1.8). Second, the average in-field DPI accuracy increased from 66.5% for one day to 75.0% for seven days, with the gain of including an additional day on accuracy dropping below 1 percentage point after 4 days.

**Conclusion:** The DPI requires ambulatory monitoring data as its discriminative power diminished significantly once computed from short in-lab recordings. Additionally, ambulatory monitoring should sample multiple days to achieve robust performance. The result of this paper can be used to make an informed decision about the tradeoff between classification accuracy and cost of data collection.

**INTRODUCTION**

Phonotraumatic vocal hyperfunction (PVH) is a subset of hyperfunctional voice disorders that is associated with the presence of phonotraumatic lesions on the vocal folds (primarily, vocal fold nodules, polyps, and reactive lesions) (Hillman et al., 2020). This subset of voice disorders is thought to be caused by accumulated trauma and mechanical stress to vocal fold tissue (Czerwonka et al., 2008; Motie-Shirazi et al., 2021; Titze et al., 2003); as such, tracking vocal function/behavior of individuals throughout their daily living can provide clinically relevant insights about factors related to the development and manifestation of this common condition. In earlier work, our group used ambulatory voice monitoring devices to investigate differences in daily vocal function/behavior between adult females with PVH and their age-, sex-, and occupation-matched vocally healthy controls (Van Stan, Mehta, Ortiz, Burns, Toles, et al., 2020). Analysis of the week-long ambulatory monitoring data showed high discriminative powers and complementary roles for standard deviation (std) of the difference between the first two harmonic magnitudes (H1-H2) and skewness of estimated sound pressure level (SPL). Logistic regression combined these two week-long distributional characteristics into a single composite measure called the Daily Phonotrauma Index (DPI) (Van Stan, Mehta, Ortiz, Burns, Toles, et al., 2020). Investigation of the components of DPI between the two groups also showed that the distribution of H1-H2 in individuals with PVH had lower std compared to controls, and that the distribution of SPL in individuals with PVH was more negatively skewed than in controls (Van Stan, Mehta, Ortiz, Burns, Toles, et al., 2020). These findings may suggest higher tendencies to use more abrupt vocal fold closures (lower H1-H2) and louder phonation (higher SPL) in individuals with PVH compared to controls during their daily living (Van Stan, Mehta, Ortiz, Burns, Toles, et al., 2020). A later study conducted a direct comparison between distribution of a wide range of ambulatory voice measures in individuals with PVH and controls and confirmed that the region with low H1-H2 was more likely to be associated with PVH, and the region with high SPL was more likely to be associated with PVH (Ghasemzadeh et al., 2024a). Later studies on DPI showed that individuals with PVH had significantly lower DPI after treatment (Van Stan et al., 2021; Van Stan, Mehta, Ortiz, Burns, Marks, et al., 2020). Thus, the DPI appears capable of providing new quantitatively-based insights into the underlying pathophysiological mechanisms associated with PVH that can be used to improve the clinical management of these disorders by playing a role in prevention (e.g., detecting increased risk for phonotrauma), diagnosis (e.g., identifying damaging patterns of voice use), and evaluating treatment outcomes (e.g., determining if treatment has reduced the risk of phonotrauma recurrence).

The current version of the DPI was trained on data collected from an entire week of ambulatory voice monitoring, which may create some challenges regarding its application. First, ambulatory voice monitoring devices are not commercially available, so the clinical utility of DPI (and any other such ambulatory measure) could be limited. One possible alternative could be to use short recordings (e.g., Rainbow passage, short spontaneous speech) that are already part of the recommended instrumental assessment of voice (Patel et al., 2018) for the computation of DPI. However, it is not known how much the discriminative power of DPI will degrade if it is computed from such short recordings. The first purpose of this study was to fill this gap by answering our first research question. **Q1**: Can DPI achieve comparable discrimination power between individuals with PVH and controls if DPI is trained on a standard, brief clinical/laboratory recording? It was hypothesized that the discriminative power of laboratory-based DPI will not be comparable to DPI based on ambulatory monitoring.

The second challenge with the application of DPI is that collecting a week of ambulatory voice monitoring data could be time-consuming, labor-intensive, and complicated. There are financial costs associated with ambulatory monitoring that will increase with the duration of monitoring. For example, in the absence of automated methods, the time needed for checking and processing the data will increase with the number of monitoring days, which will translate into an increased workload for the team and can lead to increased financial costs. Also, more monitoring days mean that the monitoring devices will rotate slower between participants, and hence, more kits will be needed, translating into the need for a larger financial investment to start a research project or clinical application of ambulatory monitoring. Reimbursement incentives to research participants may also be proportional to the number of monitoring days. Retention and participant compliance are other factors that need to be considered. Participants may start to feel uncomfortable wearing the device (e.g., skin reaction to the double-sided tape used to attach the sensor, aesthetic considerations, etc.) as the number of monitoring days is increased. Finally, analysis of our existing ambulatory monitoring data has shown that the number of participants with usable data will decrease as the threshold for the minimum number of days is increased. This means, in reality, fewer participants will be excluded from a study if the threshold for the minimum number of monitoring days is lowered, hence leading to more generalizable findings due to conducting the study on a larger sample size. In summary, it is highly desirable to reduce the number of monitoring days while maintaining the discriminative power of DPI. Thus, the second purpose of this study was to provide evidence-based recommendations for the number of monitoring days, and our second research question was **Q2**: How is the discrimination of individuals with PVH from controls based on DPI affected by the duration of ambulatory voice monitoring? It was hypothesized that ambulatory monitoring across multiple days is needed to optimize classification performance with DPI.

In earlier work, we showed that certain regions of the distribution of H1-H2 and SPL were discriminative between individuals with PVH and controls with no voice disorder (Ghasemzadeh et al., 2024a). Those results suggested that values from those regions need to be present adequately in H1-H2 and SPL sequences for their distributions to be discriminative between individuals with PVH and controls. These findings provide further support for our research questions and related hypotheses. The paper consists of two experiments, each answering one of our research questions and testing one of the related hypotheses.

**MATERIALS AND METHODS**
**Participants**

As part of an ongoing study, 134 females with PVH and 134 vocally typical female controls were prospectively enrolled for this study. Different subsets of this database have been used in earlier studies for investigating other purposes (Ghasemzadeh et al., 2024a; Mehta et al., 2015; Van Stan et al., 2023; Van Stan, Mehta, Ortiz, Burns, Toles, et al., 2020). All participants provided informed consent, and the experimental protocols were approved by the institutional review board at Mass General Brigham (protocol number: 2011P002376. Last approved 4/15/2024). Each control was closely matched to an individual with PVH based on sex, occupation, and age (±5 years). Diagnosis of the individuals with PVH was made at the Massachusetts General Hospital Voice Center based on a complete team evaluation by laryngologists and speech-language pathologists at the Massachusetts General Hospital Voice Center. The assessment included (a) the collection of case history, (b) stroboscopic imaging of the larynx, (c) auditory-perceptual evaluation using the Consensus Auditory-Perceptual Evaluation of Voice (CAPE-V) (Kempster et al., 2009) by their treating speech-language pathologist, and (d) aerodynamic and acoustic assessment of vocal function (Patel et al., 2018). Control participants had no histories of voice disorders, and their normal vocal status was verified via an interview with a licensed speech-language pathologist and laryngeal stroboscopic examination to confirm the absence of vocal fold lesions. Additionally, the participants were followed up during the duration of monitoring for any changes in their vocal function (e.g., self-reported upper respiratory infection). Under such circumstances, the monitoring was halted until the participant reported recovery and returned to normal vocal function. It must be noted that, while only obvious cases were included in the study and cases that were not clear cut were eliminated, no formal procedures were used to label the database (PVH vs. controls) that would allow for the assessment of inter- and intra-rater reliability of the labels (e.g., repeated labeling by multiple blinded clinicians).

Among the 134 individuals with PVH, 124 individuals were diagnosed with bilateral vocal fold nodules, 8 with unilateral vocal fold polyps, and 2 with each having a left vocal fold polyp and a right vocal fold nodule. The age characteristics of individuals with PVH and controls were (mean±std) 24.5±8.8 years and 24.7±8.1 years, respectively. Out of the 268 participants, 76.5% of the participants (205 individuals) identified as White, 11.2% (30 individuals) as Black or African American, 4.1% (11 individuals) as Asian, 6.7% (18 individuals) as belonging to more than one race, and 1.5% (4 individuals) did not report their race. Additionally, 90.3% (242 individuals) identified as Not Hispanic or Latino, 7.5% (20 individuals) as Hispanic or Latino, and 2.2% (6 individuals) did not report this ethnicity.

Table 1 demonstrates the occupation breakdown of individuals with PVH, where any occupation with more than one instance has been spelled out, and occupations with only one instance were grouped under the "Other" category. Table 2 presents the standard scores of the subscales and total scale for the self-reported Voice-Related Quality of Life (V-RQOL) (Hogikyan & Sethuraman, 1999). The CAPE-V scores were rated by the patient's treating speech-language pathologist and are only reported for descriptive purposes.

Table 1. Occupation breakdown of participants with PVH.

| Occupation | Count | Occupation | Count |
| --- | --- | --- | --- |
| Student | 45 | Administrator | 4 |
| Non-classical singer | 38 | Nurse | 3 |
| Classical singer | 9 | Consultant | 2 |
| School teacher | 7 | Voice teacher | 2 |
| Music teacher | 5 | Other | 19 |

Table 2. Participants self-reported V-RQOL and patients' auditory-perceptual evaluation of voice judged by a speech-language pathologist using the CAPE-V instrument.

| Measure | Subscale | PVH (mean±std) | Control (mean±std) |
|---|---|---|---|
| V-RQOL | Social-emotional | 70.2±22.6 | 95.6±7.2 |
| | Physical functioning | 69.1±21.7 | 94.9±6.3 |
| | Total score | 69.4±19.7 | 95.2±5.9 |
| CAPE-V | Overall severity | 25.7±13.9 | - |
| | Roughness | 17.7±14.0 | - |
| | Breathiness | 16.0±13.5 | - |
| | Strain | 17.6±12.6 | - |
| | Pitch | 5.7±9.9 | - |
| | Loudness | 3.1±7.7 | - |

**In-lab and in-field data collection and processing**

A set of custom-built ambulatory voice monitoring systems consisting of an accelerometer and an Android smartphone were used for data collection (Mehta et al., 2012). The accelerometer was a high-bandwidth, single-axis accelerometer (model BU-27135, Knowles Electronics, Itasca, IL, USA) and served as the main sensor for capturing phonation. Hypoallergenic double-sided tape (Model 2181, 3M, Maplewood, MN, USA) attached the accelerometer to the neck of participants halfway between the thyroid prominence and suprasternal notch (Mehta et al., 2015). The Android smartphone (Nexus 5, Google) digitized the accelerometer data at a sampling frequency of 11,025 Hz and 16-bit quantization. It also provided the hardware and software for storing the accelerometer data, real-time data processing, and interacting with the participants. The smartphone prompted the user to perform a calibration procedure using a handheld audio recorder (H1 Handy Recorder, Zoom Corporation, Tokyo, Japan) while wearing the accelerometer sensor at the beginning of every monitoring day. The calibration procedure included three repetitions of the vowel "ah" from a loud to soft phonation and allowed the estimation of SPL from the accelerometer data using a linear regression (Mehta et al., 2012). Figure 1 depicts the device.

As part of a more extensive in-lab protocol, participants read the first paragraph of the Rainbow Passage (Fairbanks, 1960), and a subset of them were prompted for approximately 30 seconds of spontaneous, monologue speech (in response to the prompt, "Tell me what you're doing today after this appointment.") while wearing the accelerometer sensor in an acoustically treated sound booth (in-lab data). After the conclusion of the in-lab session, participants were sent home with the ambulatory voice monitoring device and were instructed to wear the accelerometer sensor during their waking hours for seven days (in-field data). In-field days with a recording duration of less than 6 hours were excluded from all analyses. The selection of 6 hours as the threshold was motivated by the trade-off between the sample size of the database to allow for running robust machine learning analysis and the need to capture adequate variability in voice use throughout the day.

The raw accelerometer recordings were processed using custom MATLAB scripts. The recorded signals were first partitioned into 50 ms, non-overlapping frames. The presence/absence of voicing was determined for each frame using a voice activity detector (Mehta et al., 2015). The distributional characteristics of H1-H2 std and SPL skewness contribute to DPI (Van Stan, Mehta, Ortiz, Burns, Toles, et al., 2020) and were computed from the voiced frames of accelerometer data. H1-H2 was computed as the difference between the magnitudes of the first and second harmonics of accelerometer data over a voiced frame, expressed in dB. SPL was estimated from neck-surface acceleration magnitude (NSAM) using the linear regression model derived from the calibration task, where NSAM was computed as the logarithm of the root mean square of the accelerometer signal over a frame. Considering that skewness only depends on the shape of a distribution and is preserved during the linear transformation of calibration, NSAM skewness can replace SPL skewness in DPI. Therefore, similar to prior studies, NSAM skewness was used instead of SPL skewness for the rest of the paper (Van Stan et al., 2021, 2023; Van Stan, Mehta, Ortiz, Burns, Marks, et al., 2020). If there were multiple recording days per participant, the computed daylong NSAM skewness and H1-H2 std were averaged over all the days to derive the final features for the computation of DPI.

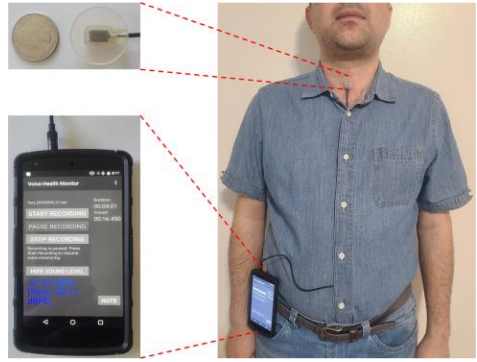

Figure 1. Image of our ambulatory voice monitoring system.

**Machine learning processing pipeline and statistical analysis**

The two components of DPI (H1-H2 std and NSAM skewness) constituted features for our machine learning analyses. Logistic regression determined the optimum decision boundary between the two classes and created the classification model. Stratified 10-fold cross-validation assessed the generalizability of the classification model. Considering that our machine-learning processing pipeline did not include any model optimization component (e.g., feature selection, hyperparameter optimization, etc.), the application of 10-fold cross-validation was appropriate (Ghasemzadeh et al., 2024b). In stratified 10-fold cross-validation, the samples from each class were randomly divided into 10 equal partitions to maintain a balanced number of patients and controls for each fold. Each partition served as the testing set of a fold, and the remaining 9 partitions served as the training set for that fold. The parameters of the machine learning model were estimated from the training set. Then, the trained model was applied to the testing set to get an estimate of its performance. Also, before training the model, features from training and testing sets separately underwent z-score normalization, where normalization parameters (mean and std of each feature) were estimated separately for each fold and only from its corresponding training samples (Ghasemzadeh et al., 2015). This separation prevented data leakage and was necessary during feature normalization.

Accuracy reflects the percentage of participants from the testing set that were classified correctly. Overall accuracy was computed as the average and std of the accuracy across the 10 testing sets. Sensitivity is the percentage of individuals with PVH from the testing set classified as individuals with PVH, and specificity is the percentage of controls from the testing set classified as controls. Overall sensitivity and specificity were computed as the average and std of their values across the 10 testing sets. The sensitivity and specificity of a classifier can be traded by changing the decision boundary of the classifier. The receiver operating characteristic (ROC) curve reflects this trade-off between a classifier's sensitivity and specificity. The area under the ROC curve (AUC) is a standard metric to quantify a ROC curve.

T-tests were used to test significant differences between the performance of in-lab and in-field DPIs, and the effect size of significant differences was quantified using Cohen's D. Spearman correlation analysis was used to quantify the association between monitoring duration and discriminative power of DPI.

**Experiment 1 materials and methods:**

All participants had in-lab Rainbow recordings. Therefore, the data from all participants (i.e., 134 females with PVH and 134 vocally typical control participants) were used for in-lab Rainbow analysis. Prompting participants for spontaneous speech was added at a later stage of the project, and only 64 patient-control pairs had such recordings. Therefore, the data from 64 pairs (i.e., 64 females with PVH and 64 vocally typical control participants) were used for in-lab spontaneous speech analysis.

This experiment had two different analyses. In the first analysis, in-lab Rainbow Passage recording was processed (please refer to In-lab and in-field data collection and processing section), and the set of in-lab Rainbow features was extracted. The in-lab Rainbow DPI was computed by training a logistic regression on in-lab Rainbow features. The same process was applied to each in-field day, and a set of daily in-field distributional characteristics was computed. Daily in-field distributional characteristics were then averaged over all days to derive the final in-field features. The in-field DPI was computed by training a logistic regression on in-field features. The second analysis followed the same steps as the first analysis, but in-lab spontaneous DPI was computed from the in-lab spontaneous speech recording.

The database for Experiment 1 was balanced (i.e., the same numbers of patients and controls) so no action was needed to make the database balanced. The only non-deterministic component of the analysis was the random splitting of the data into training and testing sets. Therefore, to make the results of machine learning analysis robust to this random effect, 10-fold cross-validation was repeated ten times with different seeds. That is, every time, different combinations of patients and controls contributed to the testing sets. The results of different repetitions of 10-fold cross-validation were pooled and then their mean and std were used for the analysis.

**Experiment 2 materials and methods:**

We needed at least seven days (each at least 6 hours in duration) of in-field monitoring to run the analyses. Forty-two individuals with PVH and 22 controls had fewer than 7 days of monitoring data and were excluded from Experiment 2. This exclusion did not preserve the matching based on occupation and age, and only matching based on sex was retained. Therefore, analyses of Experiment 2 were carried out based on the data from 92 individuals with PVH and 112 controls.

This experiment had two different analyses. In the first analysis, each in-field day recording was processed (please refer to In-lab and in-field data collection and processing section), and a set of daily in-field distributional characteristics was extracted. Different numbers of daily in-field distributional characteristics were averaged to derive the final in-field features. The number of in-field monitoring days used to compute in-field features was increased from 1 to 7, in increments of one day. The days were selected consecutively from the monitoring days, and days with a recording duration of less than 6 hours were excluded from this analysis. The in-field DPI was computed by training a logistic regression on the final in-field features.

We saw a general improvement in classification performance with more days included in the first analysis of Experiment 2. This can be interpreted in two different ways. First, more hours of recording were included in the analysis. Second, the voice use of our participants was sampled across multiple days, and hence, the DPI had better access to information related to day-to-day variations (between-day variations) of the voice use of our participants. Thus, the second analysis of Experiment 2 was conducted to tease these two components apart. The analysis was similar to the first analysis with the exception that the total duration of recording for each case was kept fixed at 6 hours (the minimum number of hours available for each day), and the number of unique days contributing to this fixed duration of 6 hours of data support was incremented from 1 to 7 days. That is, for the one-day case, a 6-hour window was selected from one day; for the two-day case, a 3-hour window was selected from each of two days; for the 3-day case, a 2-hour window was selected from each of three days, etc. The sampled day(s) and their selected window(s) were randomly selected. Since extraction of the features required the existence of voiced frames in the selected windows, another random window was chosen if the selected window had less than 6000 voiced frames (i.e., 5 minutes of voicing).

The number of patients and controls were different in Experiment 2. Hence, the dataset was balanced by an undersampling procedure (Fernández et al., 2018). The undersampling procedure randomly selected 92 controls from the 112 controls and used their data for running the 10-fold cross-validation. To make the results of the analysis robust to different random selections (i.e., the random splitting of the data into training and testing sets, the random selection of controls during the undersampling procedure, and the random selection of days and windows in the second analysis of Experiment 2), each analysis was repeated one thousand times with different seeds. That is, every time, a different random subset of controls was selected during the undersampling procedure, different combinations of patients and selected controls contributed to the testing sets, and different random day(s) and random window(s) from the recording were used in the second analysis of Experiment 2.

**RESULTS**

Two experiments were conducted in this study, each answering one of our research questions.

**Experiment 1: In-lab vs. in-field DPI**

This experiment answered the research question "**Q1**: Can DPI achieve comparable discrimination power between individuals with PVH and controls if DPI is trained on a standard, brief clinical/laboratory recording?". To that end, the classification performance of in-lab DPI models (the model trained on in-lab features) was contrasted with in-field DPI (the model trained on in-field features). The dataset was shuffled ten times to make the results robust to different data splitting, and the model's performance was evaluated each time. Table 3 contrasts the performance of in-lab and in-field DPIs. It must be noted that the difference in classification performance between the two in-field DPIs (the one in the Rainbow passage column vs. in the Spontaneous speech column) is due to differences in the participants used to run the analysis (134 pairs in Rainbow analysis vs. 64 pairs in Spontaneous analysis). The average accuracy of the in-field DPI was 15.5 percentage points higher than its in-lab Rainbow counterpart. The difference was significant with a very large effect size ($p<0.0001$, D = 1.80) (Sawilowsky, 2009). ROC curves of the two models are compared in Figure 2(A). The AUC of in-lab Rainbow DPI and in-field DPI, respectively, were 0.64 and 0.81, confirming that in-field DPI offers a better trade-off between sensitivity and specificity than in-lab DPI. The average accuracy of the in-field DPI was 24.3 percentage points higher than its in-lab spontaneous speech counterpart. The difference was significant with a very large effect size ($p<0.0001$, D = 1.87) (Sawilowsky, 2009). ROC curves of the two models are presented in Figure 2(B). The AUC of in-lab spontaneous speech and in-field models, respectively, were 0.57 and 0.81, confirming that in-field DPI offers a better trade-off between sensitivity and specificity than in-lab DPI, whether an individual reads or produces spontaneous speech.

Table 3. Comparison between the performance of in-lab and in-field DPIs. Analysis of the Rainbow recording was based on 134 patient-control pairs. Analysis of the spontaneous speech recording was based on 64 patient-control pairs.

| Model | Participants with Rainbow passage | | | Participants with spontaneous speech | | |
|---|---|---|---|---|---|---|
| | Accuracy (mean±std) | Sensitivity (mean±std) | Specificity (mean±std) | Accuracy (mean±std) | Sensitivity (mean±std) | Specificity (mean±std) |
| In-lab DPI | 57.9±10.0 | 64.4±15.1 | 51.4±14.3 | 48.9±13.3 | 57.4±21.6 | 40.4±21.8 |
| In-field DPI | 73.4±7.0 | 71.1±9.9 | 75.6±10.4 | 73.2±12.7 | 74.8±18.1 | 71.6±18.9 |

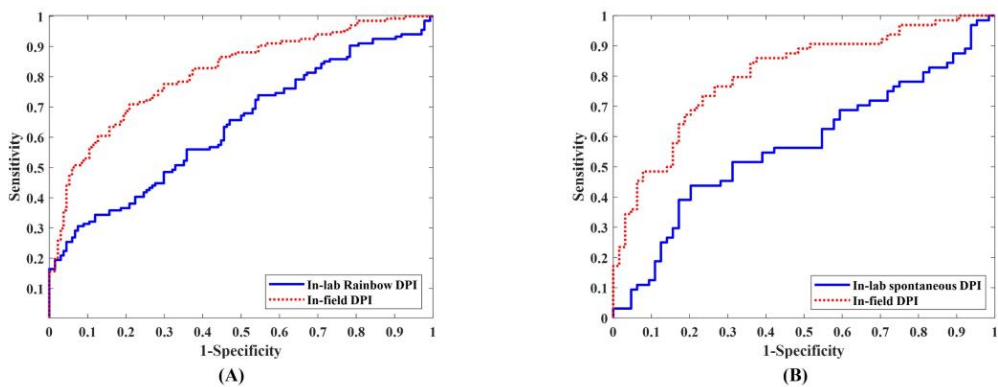

Figure 2. ROC curves comparing in-field DPI models with in-lab DPI models from the (A) Rainbow passage and (B) spontaneous speech recordings.

To further investigate the source of this significant performance gap between in-lab and in-field DPIs, differences between the two components of DPI (NSAM skewness and H1-H2 std) were compared between individuals with PVH and controls. Table 4 shows that, while in-field NSAM skewness had a large effect size between the two groups, in-lab NSAM skewness was not statistically different between the two groups during both Rainbow passage and spontaneous speech recordings. Also, while in-field H1-H2 std had a large effect size between the two groups, in-lab H1-H2 std exhibited a small-to-medium effect size between the two groups during the Rainbow passage, and H1-H2 std was not statistically different between the two groups during the spontaneous speech recording.

Table 4. Comparison between discriminative power of NSAM skewness and H1-H2 std (the two components of DPI) in the in-lab and in-field recording conditions. Analysis of the Rainbow recording was based on 134 patient-control pairs. Analysis of the spontaneous speech recording was based on 64 patient-control pairs.

**Experiment 2: The effect of the number of monitoring days**

This experiment answered the research question "**Q2**: How discrimination of individuals with PVH from controls based on DPI is affected by the duration of ambulatory voice monitoring?". To answer this question, two separate analyses were conducted.

In the first analysis, the number of in-field monitoring days used to train and compute in-field DPI was incremented from 1 to 7. The relationship between the number of monitoring days (independent variable) and the average 10-fold cross-validation accuracy of the in-field DPI (dependent variable) was non-linear (Figure 3(A)), so a Spearman correlation analysis was conducted to quantify their association. The analysis showed a strong correlation ($\rho = 0.81$, $p < 0.0001$) between the number of monitoring days and the average accuracy of in-field DPI. However, the relationship between classification accuracy and number of monitoring days was not linear and exhibited a plateau. The average accuracy of in-field DPI using only one monitoring day was 66.5 %. This number increased to 75.0 % when DPI was computed using seven monitoring days. A power function curve was fitted to the data that can be used to estimate the tradeoff between the improvement in the performance (i.e., classification accuracy) and the cost of data collection (number of monitoring days). Let $x$ denote the number of monitoring days and $y$ represent the average classification accuracy; equation 1 shows the power function:

$$y = \alpha x^b + c \qquad (1)$$

Figure 3(A) shows the estimated accuracy for including different numbers of monitoring days in the analysis, and Figure 3(B) indicates the estimated gain in the accuracy for incrementing the number of monitoring days by one day. Two example thresholds of 1 percentage point and 0.5 percentage point improvement in accuracy have also been plotted. The curve of Figure 3(A) exhibited a plateau, and the increase in the estimated accuracy dropped below 1 percentage point after including four days (versus including five days) and below 0.5 percentage point after including five days (versus including six days). Finally, the effects of including multiple days of recording on the effect size of components of DPI (NSAM skewness and H1-H2 std) were investigated. Figure 3(C) shows the result. In summary, the results of Figure 3 suggest that the robust performance of DPI depends on acquiring multiple days of recordings, in particular, to obtain discriminative power for NSAM skewness.

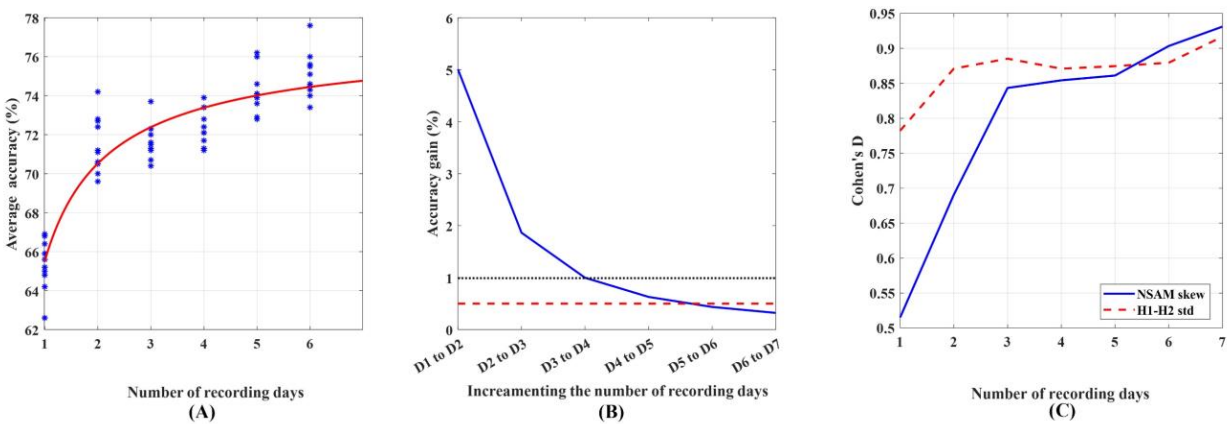

Figure 3. The effect of including different numbers of in-field monitoring days. (A) The average accuracy of in-field DPI with a power function curve fitted to data points, (B) the estimated gain in the accuracy of in-field DPI for incrementing the recording duration by one day. The red dashed line shows the threshold at 0.5%, and the black dotted line shows the threshold at 1% improvement, (C) the effect size of the two components of DPI (NSAM skewness and H1-H2 std).

The improvement observed in Figure 3 can be interpreted in two different ways. First, we included more hours of recording in the analysis. Second, we sampled the voice use of our participants over multiple days. Hence, the method had better access to information related to day-to-day variations (between-day variations) of voice use of our participants. A second analysis was conducted to tease these two components apart, where the total duration of recording for each case was kept fixed at 6 hours (the minimum number of hours available for all days), and the number of unique days contributing to this fixed duration of 6 hours of data support was incremented from 1 to 7 days. Figure 4(A) shows the result. We see improvement in the performance as the number of different monitoring days contributing to the 6-hour window increases from one to four. During this rising section of the curve (day 1 to day 4), there was a significant positive correlation between the average accuracy of the classifier and the number of monitoring days ($\rho = 0.40$, $p < 0.00001$) The correlation coefficient dropped to $\rho = 0.34$ when all 7 days were included in the analysis. In summary, the results of Figure 4(A) confirm that between-day variation of voice use is a significant source of information. To better understand the source of such information, the effects of including multiple days of recording on the effect size of the two components of DPI (NSAM skewness and H1-H2 std) were quantified using Cohen's D. Figure 4(B) shows the results, which confirms a general improvement in discriminative power of both components of DPI as more days were included in the analysis. However, it must be noted that discriminative power of NSAM skewness and H1-H2 std started to decrease after 4 and 6 days, respectively. Finally, comparing the results of Figure 4(A) with the performance of in-lab DPIs from Table 3 shows that even having multiple short in-field recordings performs much better than in-lab DPI.

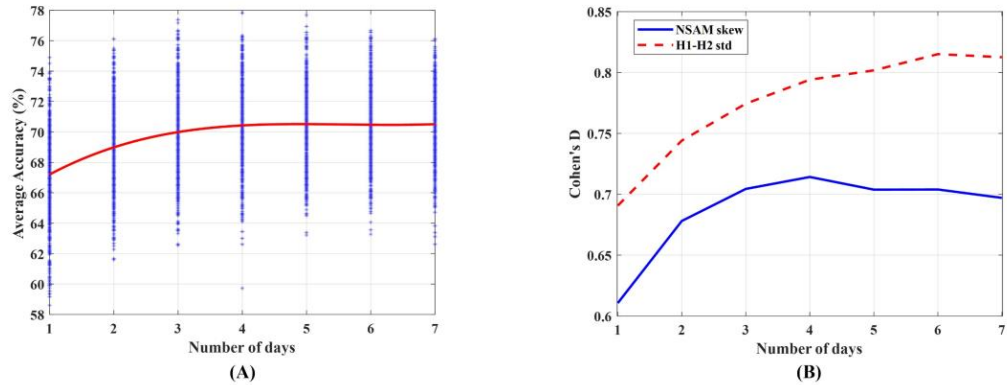

Figure 4. The effect of including different numbers of monitoring days on DPI classification accuracy while keeping the total duration of data support fixed to 6 hours. (A) The average accuracy, (B) the effect size of the two components of DPI (NSAM skewness and H1-H2 std).

## DISCUSSION

The Daily Phonotrauma Index (DPI) has provided new knowledge about pathophysiological mechanisms associated with PVH (Van Stan et al., 2021, 2023; Van Stan, Mehta, Ortiz, Burns, Marks, et al., 2020). However, the original computation of DPI depended on acquiring about a week of monitoring data, where each day of data collection can be a burden on the patient (data collection) and the clinician (data analysis and interpretation). It also requires specific hardware for data collection and specialized software for data processing, all of which may lead to additional financial expenses. So, one may ponder if ambulatory voice monitoring is necessary for accurate DPI performance, i.e., whether a comparable performance can be achieved using short in-lab recordings (e.g., Rainbow passage or spontaneous speech tokens) that are already part of the recommended instrumental assessment of voice (Patel et al., 2018).

Experiment 1's results showed that the average accuracy of the in-lab Rainbow DPI, and in-lab spontaneous DPI were 57.9% and 48.9%, respectively. To put These numbers into perspective, the one-sided 95% confidence interval of the performance of machine learning with random features (i.e., the null hypothesis) was computed following the methodology presented in our earlier works (Ghasemzadeh et al., 2024b, 2024a). Specifically, random features with the same dimensionality as in-lab features (two random features with a sample size of 134 pairs for Rainbow analysis and 64 pairs for spontaneous speech analysis) were generated 5000 times. The random features were then passed through the exact same machine learning processing pipeline presented in the Machine learning processing pipeline and statistical analysis section. The upper bound of the null hypothesis's confidence interval for Rainbow analysis was 56.0%. This number means if we use irrelevant features, we can get an average 10-fold classification accuracy of 56.0% with a probability of 5% ($\alpha = 0.05$). Therefore, the average classification accuracy of 57.9% for in-lab Rainbow DPI is not much higher than what we may have gotten if we had not collected any data and used random features instead. Similarly, the upper bound of the null hypothesis's confidence interval for spontaneous speech analysis was 58.7%, which indicates that in-lab spontaneous DPI was not performing better than random features. In contrast with in-lab DPIs, the average accuracy of in-field DPI was significantly higher at 73.4% with a very large effect size (Cohen's D = 1.8) (Sawilowsky, 2009). Therefore, DPI appears to need more voicing data than what is obtained in a short traditional in-clinic recording to perform appropriately.

Reviewing the results of Table 4 provided further insight into differences between in-lab and in-field vocal function/vocal behavior. Whereas in-field NSAM skewness had a large effect size, in-lab Rainbow NSAM skewness and in-lab spontaneous NSAM skewness did not exhibit significant discriminative power. More negative NSAM skewness in individuals with PVH is thought to align with the clinical assumption that these patients are more likely to phonate at higher vocal intensities during their activities of daily living (Ghasemzadeh et al., 2024a). This difference between individuals with PVH and controls mainly manifests in the tails of the SPL distribution (Ghasemzadeh et al., 2024a) and not in its mean (Mehta et al., 2015), suggesting that the tendency to use louder phonation in individuals with PVH compared to controls are not present at all times and may even start to emerge in certain conditions and environments. For example, we often modulate and adjust our SPL (a calibrated correlate of NSAM) based on the communication scenario that we are in. The Lombard effect (Bottalico et al., 2017; Lombard, 1911; Whittico et al., 2020) (the tendency to increase our loudness in noisy communication scenarios) and compensating for the distance between communication parties (Cheyne et al., 2009) are some examples of such adjustments. Personality trait is another factor that could indirectly affect SPL in certain conditions. For example, many individuals with PVH have been shown to be more extroverted and socially dominant than their controls, meaning they may more actively seek large social gatherings with background noise and speak over others (Roy et al., 2000). Such adjustments to SPL need specific social contexts and environments to manifest and, hence, are most likely to exhibit during in-field (ambulatory) monitoring rather than during a brief in-lab recording. In summary, the above-described adjustments to SPL occur during daily activities in response to varying vocal demands and across different communication situations. The absence of such modulators (which can only be realized through exposure to a wide range of real-life communication situations) in the lab most likely contributes to the lack of discriminative power for in-lab SPL.

The results of Table 4 showed a consistent degradation in the discriminative power of NSAM skewness for in-lab data compared to in-field data. Specifically, in-field NSAM skewness had a large effect size, whereas in-lab Rainbow and in-lab spontaneous NSAM skewness didn't exhibit significant discriminative power. Similarly, the results of Table 4 showed a consistent degradation in the discriminative power of H1-H2 std for in-lab data compared to in-field data. Specifically, in-field H1-H2 std had a large effect size, in-lab Rainbow H1-H2 std exhibited a small-to-medium effect size, and in-lab spontaneous H1-H2 std did not exhibit significant discriminative power. Investigating phonation time and percent phonation time (i.e., the ratio of phonation time to the duration of the recording) did not show significant differences between Rainbow tokens and spontaneous speech tokens and hence could not be the reason for the lack of discriminative power for in-lab spontaneous speech H1-H2 std. Differences in participants' characteristics between the two sets of in-lab analyses would also be very unlikely to explain differences between the discriminative power of in-lab Rainbow H1-H2 std and in-lab spontaneous H1-H2 std, as the discriminative power of their in-field counterparts were quite close to each other (D=0.92 vs. D=0.87). Therefore, inherent differences between reading the Rainbow passage vs. spontaneous speech should be the likely reason. Future studies may investigate more closely the factors leading to such a difference.

Once the necessity of ambulatory voice monitoring for DPI was established, we turned our attention to the practical question of the appropriate duration of monitoring. The effect of increasing the duration of monitoring from 1 to 7 days on classification performance of DPI was investigated. Our findings showed that increasing the number of monitoring days led to a general improvement in performance. This finding aligns with our earlier argument that the subtle differences in vocal function/behavior of individuals with PVH and controls may emerge more clearly as individuals are exposed to a wide range of communication situations and vocal demands. The question of which conditions or communication situations would elicit a stronger manifestation of differences in vocal function/behavior of individuals with PVH and controls is yet to be answered. Investigating variation in differences between vocal function/behavior of individuals with PVH and controls at different hours of the day and different days of the week (e.g., workdays, weekends, etc.) is a potential candidate for answering this question, which is left for a future study.

The results of Figure 3(A) can be used to estimate the expected classification performance of DPI by including different numbers of monitoring days in the analysis, and Figure 3(B) can be used to estimate the gain in performance by incrementing the number of monitoring days by one. It is worth mentioning that determining the appropriate number of monitoring days is an optimization problem that should consider the benefit and cost of increasing the number of monitoring days, where the results of Figure 3 provide the estimated benefit (i.e., improvement in the performance of DPI). The other part of the trade-off would be the cost of increasing the number of monitoring days. Such costs may include financial costs of data collection (e.g., paying incentives to participants, processing more data, slower rotations of monitoring kits between participants and hence the need for a greater number of kits, etc.), retention of participants (e.g., participants may opt out of the study if they have to wear the device for many days), sample size of the database (participants with monitoring days fewer than the threshold will be excluded hence reducing the sample size of the final database), and other factors. Considering that the cost of increasing the number of monitoring days was not modeled and quantified in the study, a precise and well-justified answer to the optimum number of monitoring days cannot be provided here. Therefore, the decision would be somewhat subjective and depends on what one considers an acceptable threshold for gain in accuracy, which is a common practice in the machine learning community (Figueroa et al., 2012). Figure 3(B) shows examples of two such thresholds chosen for illustrative purposes. If the threshold of 1 percentage point is selected, 4 days of monitoring would be adequate, and if the threshold of 0.5 percentage point is selected 5 days of monitoring would be adequate. Last but not least, Figure 3(A) shows the plot exhibits a plateau, which has been reported in many machine learning studies investigating the effect of sample size and data support (Figueroa et al., 2012; Ghasemzadeh et al., 2024b; Viering & Loog, 2022); therefore, it seems unlikely that extending the number of monitoring days beyond 7 days (at least in our population) would provide significant improvement.

The result of Figure 4(A) suggested that between-day variation of voice use could be a significant source of information. However, the drop in the correlation coefficient and saturation in the plot after four days seems to be at odds with the findings of Figure 3(A) and requires further explanation. In the analysis of Figure 4(A), as the number of days was increased, the duration of data used from each day was reduced. This means access to information related to the within-day variation of voice use between individuals with PVH and controls was limited for cases with more days. Therefore, both within-day and between-day variations contribute novel information to the analysis, and participants need to be monitored adequately across different days and during each day. To further study this finding, the effect of including different numbers of monitoring days on the discriminative power of the two components of DPI was investigated in both analyses of Experiment 2 (Figures 3(C) and 4(B)). One interesting finding was that the effect size of NSAM skewness increased from $D = 0.52$ (for one day) to $D=0.93$ (for 7 days), exhibiting the biggest improvement in the discriminative power of the components of DPI (Figure 3(C)). This result agrees with our earlier arguments regarding SPL and the necessity of monitoring participants subjected to varying vocal demands and communication situations during activities of daily living. Additionally, the result of Figure 4(B) indicates a consistent improvement in the discriminative power of H1-H2 std as the number of monitoring days is increased, whereas the discriminative power of NSAM skewness dropped after including 4 days. This suggests adequate monitoring during each day (within-day variation) is more critical for NSAM skewness compared to H1-H2 std. In a recent study, we found that NSAM skewness contributed more to classification of individuals with mild PVH than H1-H2 std (Van Stan et al., 2023). Therefore, we may hypothesize that more data is needed for robust classification of individuals with mild PVH compared to individuals with moderate/severe PVH. This is a direction that a future study can investigate.

It must be noted that the purpose of ambulatory monitoring in general, and DPI specifically, is not to replace the routine clinical evaluation but rather to augment it by providing new insights into the daily vocal function/behavior of individuals with PVH that can be used to improve the clinical management of these disorders (e.g., prevention, more ecologically valid assessment of treatment outcomes, etc.). Therefore, the low classification accuracy of the in-lab DPI should not be interpreted as the limited discriminative power of procedures that are employed in the lab or clinic. Such procedures include stroboscopy which was actually used as a primary factor in the ground truth labeling of participants as PVH vs. controls. The low classification accuracy of in-lab DPI means most of the discriminative information that was present in the in-field data was not present in the in-lab data. Therefore, the information obtained from DPI (and ambulatory monitoring studies in general) cannot be gained from in-lab data, which highlights the complementary role of ambulatory monitoring.

Despite the importance of our current findings, several limitations should be mentioned. Like any other study, our findings' generalizability is bounded by our participants' demographic characteristics. All of our participants were female, and most were young, with mild severity of dysphonia (Table 2), and affiliated with the singing community (Table 1). These characteristics should be considered when generalizing the results of this study to other populations. In addition, our selection of the variable of interest was limited by the existing literature and the measures that have been proposed so far. Therefore, we only investigated a limited set of acoustic variables (a linear combination of standard deviation of H1-H2 distribution and skewness of NSAM distribution) in our analyses.

**CONCLUSION**

The DPI is a recently validated measure of PVH, but it requires effort and resources for data collection and processing, which could be costly and time-consuming. The purpose of this study was to see if such costs could be reduced. The results of the first experiment of this study showed that DPIs computed from short clinical recordings (e.g., Rainbow passage, short spontaneous speech) that are already part of the recommended instrumental assessment of voice do not offer acceptable discriminative power. Therefore, pathophysiological mechanisms associated with daily voice use in individuals with PVH, as quantified by DPI, cannot be gained from in-lab data, which highlights the complementary role of ambulatory monitoring. The results of the second experiment showed a general improvement in the accuracy of DPI, as well as the discriminative power of the two components of DPI for increasing the duration of monitoring. Interestingly, SPL skewness benefited the most from multiple days of monitoring. The result of this analysis can be used to make an informed decision about the tradeoff between the performance (i.e., classification accuracy) and the cost of data collection (the number of monitoring days). Classification accuracy plateaued after including four to five days of monitoring, indicating a sweet spot for such a decision.

**Acknowledgments**

This research was funded by the National Institutes of Health - National Institute on Deafness and Other Communication Disorders (grants: K99 DC021235, T32 DC013017, R33 DC011588, P50 DC015446). The article's contents are solely the responsibility of the authors and do not necessarily represent the official views of the NIH. The authors would like to thank Laura Toles and Katie Marks for study participant recruitment and data collection, MGH Voice Center research assistants for data pre-processing, AJ Ortiz for data analysis and signal processing, and Rob Petit and Dave Viggiano for mobile application development.

**Data availability statement**

**REFERENCES**

Bottalico, P., Passione, I. I., Graetzer, S., & Hunter, E. J. (2017). Evaluation of the starting point of the Lombard effect. *Acta Acustica United With Acustica*, *103*(1), 169–172.

Cheyne, H. A., Kalgaonkar, K., Clements, M., & Zurek, P. (2009). Talker-to-listener distance effects on speech production and perception. *The Journal of the Acoustical Society of America*, *126*(4), 2052–2060.

Czerwonka, L., Jiang, J. J., & Tao, C. (2008). Vocal nodules and edema may be due to vibration-induced rises in capillary pressure. *The Laryngoscope*, *118*(4), 748–752.

Fairbanks, G. (1960). *Voice and articulation drillbook* (Vol. 127). Harper New York.

Fernández, A., Garcia, S., Galar, M., Prati, R. C., Krawczyk, B., & Herrera, F. (2018). *Learning from imbalanced data sets* (Vol. 10, Issue 2018). Springer.

Figueroa, R. L., Zeng-Treitler, Q., Kandula, S., & Ngo, L. H. (2012). Predicting sample size required for classification performance. *BMC Medical Informatics and Decision Making*, *12*, 1–10.

Ghasemzadeh, H., Hillman, R. E., & Mehta, D. D. (2024a). Consistency of the Signature of Phonotraumatic Vocal Hyperfunction Across Different Ambulatory Voice Measures. *Journal of Speech, Language, and Hearing Research (Accepted)*.

Ghasemzadeh, H., Hillman, R. E., & Mehta, D. D. (2024b). Toward Generalizable Machine Learning Models in Speech, Language, and Hearing Sciences: Estimating Sample Size and Reducing Overfitting. *Journal of Speech, Language, and Hearing Research*, *67*(3), 753–781.

Ghasemzadeh, H., Tajik Khass, M., Khalil Arjmandi, M., & Pooyan, M. (2015). Detection of vocal disorders based on phase space parameters and Lyapunov spectrum. *Biomedical Signal Processing and Control*, *22*, 135–145. https://doi.org/10.1016/j.bspc.2015.07.002

Hillman, R. E., Stepp, C. E., Van Stan, J. H., Zañartu, M., & Mehta, D. D. (2020). An updated theoretical framework for vocal hyperfunction. *American Journal of Speech-Language Pathology*, *29*(4), 2254–2260.

Hogikyan, N. D., & Sethuraman, G. (1999). Validation of an instrument to measure voice-related quality of life (V-RQOL). *Journal of Voice*, *13*(4), 557–569.

Kempster, G. B., Gerratt, B. R., Abbott, K. V., Barkmeier-Kraemer, J., & Hillman, R. E. (2009). Consensus auditory-perceptual evaluation of voice: development of a standardized clinical protocol. *American Journal of Speech-Language Pathology*, *18*(2), 124–132.

Lombard, E. (1911). Le signe de l'élévation de la voix (translated from French). *Ann. Des Mal. l'oreille Du Larynx*, *37*(2), 101–119.

Mehta, D. D., Van Stan, J. H., Zañartu, M., Ghassemi, M., Guttag, J. V, Espinoza, V. M., Cortés, J. P., Cheyne, H. A., & Hillman, R. E. (2015). Using ambulatory voice monitoring to investigate common voice disorders: Research update. *Frontiers in Bioengineering and Biotechnology*, *3*, 155.

Mehta, D. D., Zanartu, M., Feng, S. W., Cheyne II, H. A., & Hillman, R. E. (2012). Mobile voice health monitoring using a wearable accelerometer sensor and a smartphone platform. *IEEE Transactions on Biomedical Engineering*, *59*(11), 3090–3096.

Motie-Shirazi, M., Zañartu, M., Peterson, S. D., & Erath, B. D. (2021). Vocal fold dynamics in a synthetic self-oscillating model: Contact pressure and dissipated-energy dose. *The Journal of the Acoustical Society of America*, *150*(1), 478–489.

Patel, R. R., Eadie, T., Paul, D., Hillman, R., Barkmeier-Kraemer, J., Awan, S. N., Courey, M., Deliyski, D., & Švec, J. G. (2018). Recommended Protocols for Instrumental Assessment of Voice: American Speech-Language-Hearing Association Expert Panel to Develop a Protocol for Instrumental Assessment of Vocal Function. *American Journal of Speech-Language Pathology*, *27*(3), 887–905. https://doi.org/10.1044/2018_ajslp-17-0009

Roy, N., Bless, D. M., & Heisey, D. (2000). Personality and voice disorders: a multitrait-multidisorder analysis. *Journal of Voice*, *14*(4), 521–548.

Sawilowsky, S. S. (2009). New effect size rules of thumb. *Journal of Modern Applied Statistical Methods*, *8*, 597–599.

Titze, I. R., Svec, J. G., & Popolo, P. S. (2003). Vocal dose measures: Quantifying Accumulated Vibration Exposure in Vocal Fold Tissues. *Journal of Speech, Language, and Hearing Research*, *44*, 919–932.

Van Stan, J. H., Burns, J., Hron, T., Zeitels, S., Panuganti, B. A., Purnell, P. R., Mehta, D. D., Hillman, R. E., & Ghasemzadeh, H. (2023). Detecting Mild Phonotrauma in Daily Life. *The Laryngoscope*, *133*(11), 3094–3099.

Van Stan, J. H., Mehta, D. D., Ortiz, A. J., Burns, J. A., Marks, K. L., Toles, L. E., Stadelman-Cohen, T., Krusemark, C., Muise, J., Hron, T., & others. (2020). Changes in a Daily Phonotrauma Index after laryngeal surgery and voice therapy: Implications for the role of daily voice use in the etiology and pathophysiology of phonotraumatic vocal hyperfunction. *Journal of Speech, Language, and Hearing Research*, *63*(12), 3934–3944.

Van Stan, J. H., Mehta, D. D., Ortiz, A. J., Burns, J. A., Toles, L. E., Marks, K. L., Vangel, M., Hron, T., Zeitels, S., & Hillman, R. E. (2020). Differences in weeklong ambulatory vocal behavior between female patients with phonotraumatic lesions and matched controls. *Journal of Speech, Language, and Hearing Research*, *63*(2), 372–384.

Van Stan, J. H., Ortiz, A. J., Marks, K. L., Toles, L. E., Mehta, D. D., Burns, J. A., Hron, T., Stadelman-Cohen, T., Krusemark, C., Muise, J., & others. (2021). Changes in the Daily Phonotrauma Index Following the Use of Voice Therapy as the Sole Treatment for Phonotraumatic Vocal Hyperfunction in Females. *Journal of Speech, Language, and Hearing Research*, *64*(9), 3446–3455.

Viering, T., & Loog, M. (2022). The shape of learning curves: a review. *IEEE Transactions on Pattern Analysis and Machine Intelligence*.

Whittico, T. H., Ortiz, A. J., Marks, K. L., Toles, L. E., Van Stan, J. H., Hillman, R. E., & Mehta, D. D. (2020). Ambulatory monitoring of Lombard-related vocal characteristics in vocally healthy female speakers. *The Journal of the Acoustical Society of America*, *147*(6), EL552–EL558.